# Overview on electrical issues faced during the SPIDER experimental campaigns

A. Maistrello[a], M. Agostini[a,b], M. Bigi[a], M. Brombin[a,b], M. Dan[a], R. Casagrande[a], M. De Nardi[a,d], A. Ferro[a], E. Gaio[a,b], P. Jain[a], F. Lunardon[a,c], N. Marconato[a,e], D. Marcuzzi[a], M. Recchia[a,b], T. Patton[a], M. Pavei[a], F. Santoro[a,d], V. Toigo[a,b], L. Zanotto[a], M. Barbisan[a], L. Baseggio[a], M. Bernardi[a], G. Berton[a], M. Boldrin[a], S. Dal Bello[a], D. Fasolo[a,e], L. Franchin[a], R. Ghiraldelli[a], L. Grando[a], R. Milazzo[a], A. Pimazzoni[a], A. Rigoni[a], E. Sartori[a,e], G. Serianni[a,b], A. Shepherd[a], M. Ugoletti[a], B. Zaniol[a], D. Zella[a,e], E. Zerbetto[a], H. Decamps[f], C. Rotti[f], P. Veltri[f]

[a]*Consorzio RFX (CNR, ENEA, INFN, Università di Padova, Acciaierie Venete SpA), Corso Stati Uniti 4, 35127 Padova, Italy*
[b]*CNR- ISTP Padova, Italy*
[c]*Department of Applied Physics, Ghent University, 9000 Gent, Belgium*
[d]*CRF – University of Padova, Italy*
[e]*Dept. of Electrical Engineering, University of Padova, Padova, Italy*
[f]*ITER Organization, Route de Vinon-sur-Verdon, CS 90 046, 13067 St Paul Lez Durance Cedex, France*

**Abstract**

SPIDER is the full-scale prototype of the ion source of the ITER Heating Neutral Beam Injector, where negative ions of Hydrogen or Deuterium are produced by a RF generated plasma and accelerated with a set of grids up to ~100 keV. SPIDER Beam Source design complies with the ITER specific requirements, being fully installed within a vacuum vessel, and with the electrical circuits operating at the residual background pressure. The Power Supply System is composed of high voltage dc power supplies capable of handling frequent grid breakdowns, high current dc generators for the magnetic filter field and RF generators for the plasma generation.

During the first 3 years of SPIDER operation different electrical issues were discovered, understood and addressed thanks to deep analyses of the experimental results supported by modelling activities.

One of the main issues encountered was the presence of RF discharges on the RF circuit on the backside of the beam source, limiting the operating pressure within the source.

The self-excited RF oscillators showed frequency instabilities that were found to be intrinsic limits of the application of this technology to the resonant loads of Neutral Beam Injectors; their understanding allowed assessing the technical basis for the final decision to replace the RF generators in SPIDER and MITICA and change the current ITER baseline to solid-state amplifier technology.

Other issues faced regard the effect of the mutual coupling between the RF circuits on board the source, various electromagnetic compatibility problems, the limitation in the operational space to be mitigated improving the SPIDER operating strategies, the revision of the configuration of the magnetic filter field layout to avoid plasma quench.

The paper gives an overview on the observed phenomena and relevant analyses to understand them, on the effectiveness of the short-term modifications provided to SPIDER to face the encountered issues and on the design principle of long-term solutions to be introduced during the currently ongoing long shutdown.

## 1. Introduction

SPIDER started its operation in June 2018 and in late 2021 started a long shutdown for improvements [1].

In SPIDER the beam source is fully installed within a vacuum vessel and operated in the residual gas pressure, including 8 radiofrequency (RF) inductively coupled plasma drivers, their matching networks, and the set of grids to extract and accelerate the negative ions at ~100 keV.

With reference to Figure 1, the power supply system is composed of the Acceleration Grid Power Supply (AGPS) [2] which biases up to -96 kV the Extraction Grid (EG) and the Ion Source and Extraction Power Supplies (ISEPS) with respect to the Grounded Grid (GG). ISEPS, installed within a Faraday cage called "High Voltage Deck" (HVD) [3], is in turn composed of [4]:

- the Ion Source Extraction Grid (ISEG) power supply biases the Plasma Grid (PG) with respect to the EG up to -12 kV;
- the Ion Source Plasma Grid (ISPG) power supply feeds a current up to 5 kA to the Plasma Grid (PG);
- the Ion Source BIas (ISBI) power supply biases the source body with respect to the PG up to -30 V;
- the Ion Source Bias Plate (ISBP) power supply biases the Bias Plate (BP) with respect to the source body up to 30 V.
- 4 RF free running generators rated for 200 kW on 50 Ω in the frequency range 900 kHz-1100 kHz, supply 4 driver pairs for the plasma generation.

The power supplies are connected to the Beam Source through a multi-conductor transmission line 40 m long [5].

SPIDER power supply system and the beam source were designed taking into account normal and anomalous

---

operation, with the support of simulation tools and benefiting from the experimental results obtained with the most significant devices. Nevertheless, the need to satisfy the specific ITER requirements and the relevant complexity of the circuits led to explore novel conditions and encounter electrical issues during the SPIDER experimental campaigns, not easily predictable.

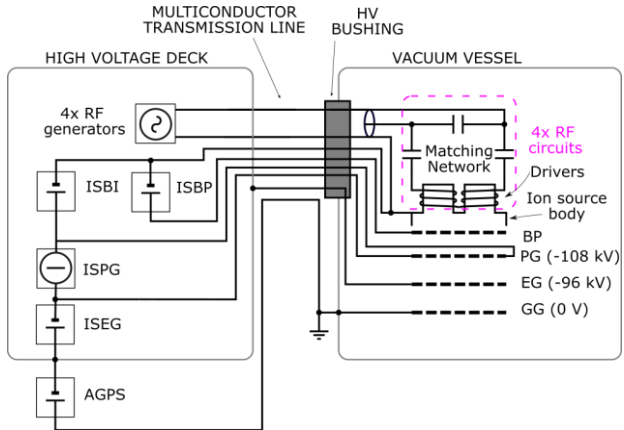

**Figure 1 – Simplified SPIDER electrical scheme**

## 2. RF system

SPIDER RF system is composed of 4 RF oscillators rated for 200 kW on a 50 Ω load, supplied by a common HV power supply rated for 12 kV and 140 A.

The RF generator is a self-excited oscillator composed of two tetrodes in push-pull configuration, a tank circuit realized with an output transformer and two variable capacitors ($C_V$) to adjust the frequency in range 900 kHz – 1100 kHz.

The simplified scheme of an oscillator is shown for reference in Figure 2.

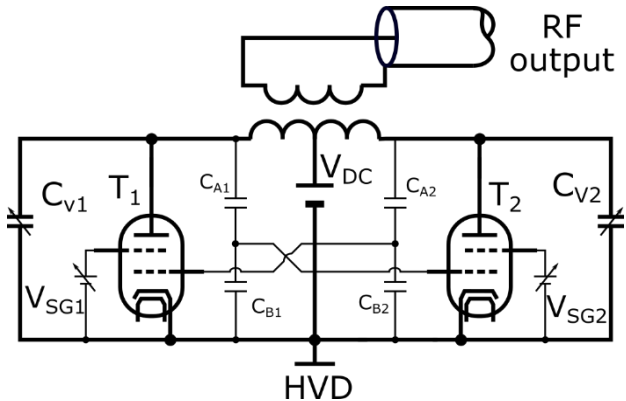

**Figure 2 – Simplified scheme of an oscillator**

In SPIDER each RF oscillator supplies a driver pair, mounted on board the beam source. The driver is a highly engineered inductively coupled plasma source whose equivalent impedance change after plasma initiation and with plasma parameters. The driver pairs parameters estimated during SPIDER operation are listed in Table 1: those for the vacuum case (without plasma) derive both from measurements with impedance meter and from the generators output probes, while the parameters with plasma are obtained from generator probes only. The equivalent inductance Ld includes the stray inductance due to the connections.

**Table 1 – Driver pair equivalent electrical parameter in vacuum condition and with plasma**

| Equivalent parameter | Vacuum | Plasma |
|---|---|---|
| Rd | 2 Ω | 3 - 5 Ω |
| Ld | 20 µH | 19.5 µH |

Each pair is matched to the generator characteristic impedance by means of a normal L-type matching network composed of a parallel capacitance of 10 nF given by two 5 nF parallel capacitors and two series capacitors each of 3 nF [6], as sketched in Figure 3.

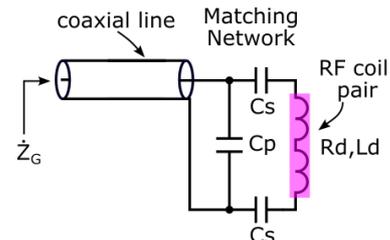

**Figure 3 – Scheme of one RF circuit of SPIDER. Rd is the stray resistance of the RF coil pair.**

The impedance seen at the output generator ($Z_G$) and the related VSWR as a function of the frequency are shown in Figure 4: top plot shows the impedance modulus, central plot shows the impedance phase displacement, and the bottom plot shows the related VSWR (vertical lines correspond to the frequency at which the VSWR is minimum, while the horizontal line is a reference for the reader that corresponds to VSWR = 2). The variation of Ld and Rd after plasma initiation and during the pulse modifies the impedance seen by the generator and the minimum VSWR. The circuit is matched when the impedance seen by the generator is ~50+j0 Ω, which corresponds to VSWR ~1. This happens only for some of the equivalent driver parameters and on a narrow bandwidth, exploiting resonances of the driver's impedance with the fixed matching capacitance.

It is highlighted that the presence of a resonance means that the circuit is subjected to RF high voltage and currents, that for SPIDER are expected up to 17 kV rms and 260 A rms at full power.

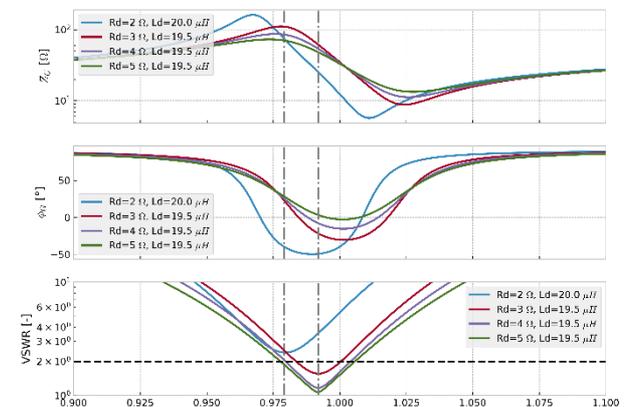

**Figure 4 – Numerical simulation of the impedance seen by the RF generator for different driver equivalent**

parameters. Top plot: impedance modulus, central plot: impedance phase displacement, bottom plot: VSWR.

### 2.1. RF oscillators frequency instability

Self-excited oscillators had been preferred since they were supposed to be capable of adaptation against the load variations; on the contrary their operation revealed one of the main issues encountered during the first SPIDER experimental campaigns: the frequency instability called "frequency flip", which is the sudden variation of the frequency observed when varying the internal variable capacitance to look for the matching condition. This phenomenon was observed also on other RF ion sources supplied with oscillators [7]. The frequency flip creates an inhibited frequency range that includes the minimum VSWR. The main consequence is that the matching condition (minimum VSWR) is not reachable and thus, the oscillator cannot deliver the rated power to the load.

The developed models allowed to deepen the understanding of the observed phenomena, which is explained by the presence of two resonant circuits, decoupled by means of an impedance: one is the tank circuit of the generator, the other one is the load, and the decoupling impedance is the transmission line [8].

The frequency flip shows a hysteretic behavior in the plane $C_V$-frequency, as shown in the bottom graph of Figure 5: starting with a low $C_V$ and increasing it the frequency reduces accordingly, but when $C_V$ reach the frequency flip low threshold the frequency jumps to a value tens of kHz lower. Further increasing $C_V$ causes a reduction in frequency. Coming back to lower $C_V$ the frequency flip threshold is different. As anticipated the best matching frequency is within the inhibited range, as sketched in the top graph of Figure 5.

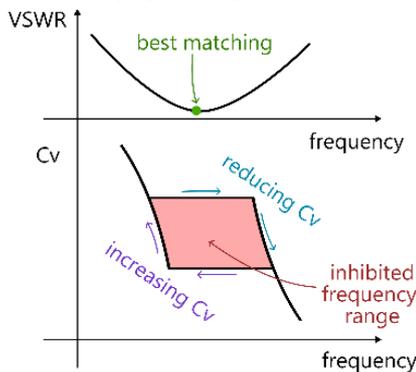

**Figure 5 – Top graph sketch of the VSWR as a function of the frequency, bottom graph sketch of the inhibited frequency range in the plane $C_V$, frequency**

The amplitude of the inhibited frequency range is directly proportional to the driver pair equivalent inductance and negatively proportional to the driver pair equivalent resistance.

Due to the change of the impedance seen by the generator after plasma initiation and with plasma parameters, the operating frequency has to be varied to reach the best achievable matching. The first strategy developed was to start the oscillator at low power and with a $C_V$ such that the frequency is close to the frequency flip low threshold; then, after plasma initiation, reduce the $C_V$ to reach the new threshold (that changed accordingly to the load variation) and afterward increase the power up to the maximum value reachable. However, in this way the cross-talk between oscillators was significant, and required the development of a different strategy, as described in section 2.2

### 2.2. RF oscillators cross-talk

The impedance seen by the RF generators is slightly different for the four RF circuits and thus the associated inhibited frequency range are different. This means that each generator has a different best working frequency without frequency flip occurrence, which was individually found experimentally with the strategy explained in section 2.1.

The operation of two or more generators together was not straightforward since a strong modulation of the output power and of the Hα appeared, due to the cross-talk. The modulation frequency is equal to the difference between the generators frequencies and was found stronger on the two top driver pairs and on the two bottom pairs, while it was weaker between the two central driver pairs. This suggested to look for differences among the driver pairs; a difference was found on the layout of the RF circuit, as sketched in Figure 6 (the matching capacitors are not shown for the sake of simplicity).

In particular the path of the RF circuit supplying pair 1 creates a large loop, that even crosses pair 2, the same states for pair 3 and 4 while this is not the case for pairs 2 and 3. The flux linkage with neighbor driver pairs due to these large loops causes mutual coupling between pairs 1-2 and pairs 3-4, which correlates to the pairs affected by large RF power and Hα modulation.

A relatively easy modification of the layout, as indicated in Figure 6 with the red arrows, was introduced during a short SPIDER shutdown. Part of the conductors were substituted such that the path laid on the plane of the driver coil winding being supplied, avoiding the large loops over a different driver pair.

The mutual inductance between driver pairs 1-2 and 3-4, extrapolated from the scatter parameters measured by means of a network analyzer, was reduced to 25% of the original value. Furthermore, the voltage induced at the coupled generators was reduced to 40% of the original value; details of the study can be found in [9].

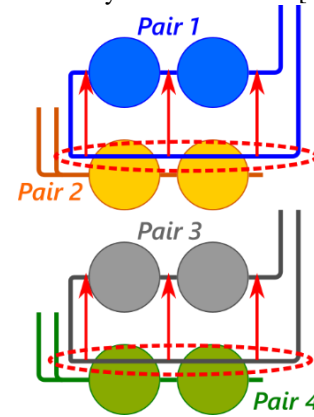

**Figure 6 – Sketch of the layout of the RF circuits on the backside of the beam source (circles represent the drivers)**

The output power and Hα modulation is still present in some operating conditions (magnetic filter field, gas pressure, RF power and frequency, …) suggesting that the coupling due to the layout was not the only root cause.

To try to avoid the cross-talk effects, specific tests were performed over a wide range of SPIDER parameters: these allowed to observe that the plasma light modulation strongly reduced when the oscillators were locked in frequency or when the frequency difference was sufficiently high. The first condition was not easily achievable and maintainable for different plasma parameters (strongly influenced by gas pressure, magnetic filter field, RF power); the latter was instead more easily obtainable even if with the drawback of a further reduction of the power delivered to the load [10]. The RF generators $C_V$ control strategy was then improved accordingly.

### 2.3. Tetrodes high voltage needs

The tetrode technology implemented on the RF oscillators requires the use of high dc voltage for the operation, which is applied also to other components such as the tank circuit and the feedback circuits. Frequently programmed maintenance for components cleaning to avoid flashover and conditioning with an external power supply to recover the variable capacitors voltage hold-off is mandatory.

For MITICA these activities are not trivial, since the oscillators are placed within a high voltage deck which sits on insulators 6 m high, with possible access only trough moving platform or temporary staircase placed by means of an overhead crane. Nevertheless, the experience with SPIDER demonstrates that preventive maintenance is not sufficient: during experimental campaigns variable capacitors required additional conditioning procedures to recover from arcs. The poor reliability of variable capacitors negatively impacted on the available experimental time, and it will be more important for MITICA given the limited accessibility.

### 2.4. RF stray current

When operating the RF generators two unwanted phenomena where observed: the circulation of a significant RF current on some plant components not intended for that RF currents, and noise on some diagnostics that overwhelmed the useful signal.

Two connection points to HVD of the RF circuit were identified: the first given by the connection of the coaxial cable outer conductor to the beam source body and to the HVD through ISBI, ISPG and ISEG; the second given by a voltage divider installed within the RF generators. The first connection point cannot be removed for functional reasons (to avoid overvoltage at grid breakdown, coaxial line EM field irradiation, …). The second connection to the HVD is the central point of a differential Capacitive Voltage Divider (CVD) provided by the oscillator' supplier and connected downstream the output transformer.

In order to reduce the stray current amplitude it was necessary to achieve a circuit downstream of the output transformer with only one point connected to the HVD; the best option was to shift the CVD to the primary side of the output transformer, assuring the necessary adaptation to maintain the expected voltage signal in amplitude and phase. A conceptual scheme is sketched in Figure 7. In this way, the output transformer stray capacitance existing between the windings, which is in the order of ~10 pF, results in series for the RF stray current path.

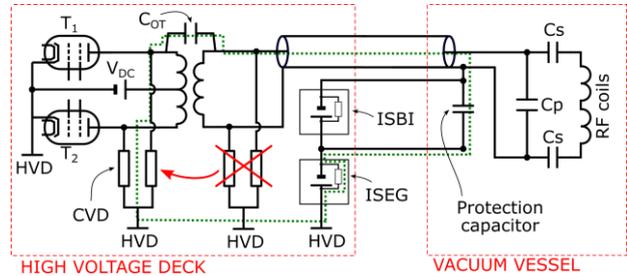

**Figure 7** – Simplified electrical circuit of an oscillator feeding a driver pair; the shift of the CVD connection points is indicated. Green dotted line indicates the path for the RF stray current after the improvement

After the modification the stray current was reduced to about 0.2 A with a single generator supplying 100 kW to the plasma source; the value is 12% to 25% of the original [9], depending on the RF generator. The noise level on some of the diagnostics affected by the RF generators operation was reduced to the normal background noise. Other root causes for the noise still affecting some other diagnostics have to be investigated.

### 2.5. Alternative technology for the RF generators: solid-state switching amplifiers

The frequency instability inherent to the RF self-excited oscillators used to supply the SPIDER RF source, that limit the maximum output power delivered to the load, cannot be fully overcome with the conceived operational strategies and improvements described in section 2.2. A more radical upgrade was deemed necessary, consisting in the changing of the technology to amplifiers. Amplifiers do not exploit internal resonant circuit, instead the frequency is directly set externally and thus the best matching frequency can be reached.

Amplifiers can all be driven at the same frequency, limiting the effects of cross-talk between the generators.

The use of solid-state technology introduces additional benefits since the components do not use high voltage: there is no need for conditioning procedures or frequent cleaning, the risk of obsolescence of tetrodes is avoided and there is no need for an additional large HV power supply. Moreover, to achieve large RF rated power a modular approach is mandatory, which means that in principle a single module failure do not stop the generator; therefore the experimental campaign can continue with limited output power that result in an increased plant availability.

Switching technology allows increasing the efficiency with respect to linear amplifier, even if the obtained square wave at the output has to be filtered-out to obtain a sinusoidal wave with limited THD.

These expected advantages pushed for the replacement of RF oscillator technology with RF solid-state switching amplifiers (RFSSA) for SPIDER and MITICA and to change the baseline for ITER HNBI;

more details can be found in [11]. In SPIDER, RF amplifiers will be installed during the long shutdown, ongoing at the time of writing.

The operative experience with RF oscillators made us aware of the importance of the care to be taken on the integration of RFSSA in SPIDER plant, where stray parameters and secondary level details (for instance transducers connections) may impact negatively on the SPIDER operational behavior. Furthermore, there is limited experience within the NBI community of RF driven ion sources supplied with RF amplifiers at high power, although positive feedbacks are being produced [12] [13].

Integration studies were performed to estimate the amplitude of stray currents in the new circuits with RFSSA, the mutual coupling between RF circuits on the generator output modulation, the impact of harmonic distortion on transmission line and RF load; the details can be found in [14].

The technical specifications for the RFSSA addresses the need to test them in factory on a load relevant for the application in normal and abnormal conditions, as close as possible to the final ones; the design studies of a suitable dummy load to this purpose and of the special tests conceived to demonstrate the design and operation of the RFSSA first unit are described in [15].

## 3. RF discharges on the beam source backside

RF induced discharges at the beam source backside with vessel pressure higher than 40 mPa were observed [16]. By looking at visible cameras it was possible to define two macro-type of discharges studied in [17]: a "glow" type where the background gas is ionized and a diffuse light is observed close to the drivers (Figure 8 a), and an "arc" type where a flashover is visible on specific areas (Figure 8 b). The transition from glow to arc type was also observed.

Signals from light detectors on the backside of the beam source and measured output current and voltage from ISBI confirm different type of phenomena. The basic understanding of the events allowed to develop a methodology to identify them in real-time by looking at the electrical measurements trends, with the aim of preventing damages to the beam source components by actuating a fast stop of the pulse.

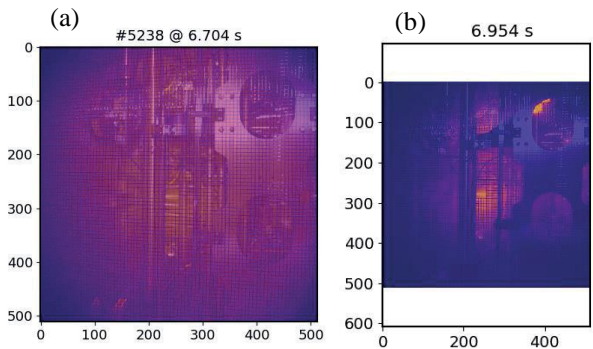

**Figure 8 – Pictures of a glow discharge (a) and of an arc discharge (b) [17]**

To proceed with the experimental campaign without incurring into a discharge it was necessary to reduce the gas conductance between the ion source and the vessel. The target conductance allowed operating the ion source at the nominal pressure, with a background pressure lower than 40 mPa. The adopted temporary provision was the installation of a PG mask that closes most of the apertures; the drawback is the limitation of the beam current. The long term solution is to increase the pumping speed [1].

Inspections during short term shutdowns and deep analyses of the RF circuit geometry including passive components at beam source potential allowed identifying locations characterized by high electric field (drivers, matching network capacitors, ceramic breaks), that can be improved to reduce the electrons emission and thus the probability to initiate a discharge.

Actually, the drivers were known being critical components for the voltage hold-off given the experience in ELISE at IPP Garching, that shared the same driver design concept [18] [19], and that was hindered by discharges in the driver region [20].

Figure 9 shows the cross section of a driver: the RF coil is tightly wound around an alumina driver case (RF window), with dielectric combs for inter-turns spacing. A faraday screen at beam source body potential protects the alumina from plasma etching. This configuration makes the electric field at the triple junction (top turn, driver case, vacuum as indicated in Figure 9 a) particularly high, allowing electron emission and thus reducing the voltage hold off.

Improvements of the driver configurations foresees the radial support of the coil from the outside, by means of 4 combs 90° displaced, introducing a gap between the coil and the driver case, as described in detail in [21]: in this way the triple junction is moved to the comb as shown in Figure 9 b, where the electric field is one order of magnitude lower.

A further improvement is obtained with the adoption of holes on the comb to allow electric field to relax close to the high voltage turn. For the matching capacitors and ceramic breaks additional shields were designed with the aim to reduce the electric field at their triple junction; the analyses and implementation of these upgrades on RF circuits on board the source are reported in [22].

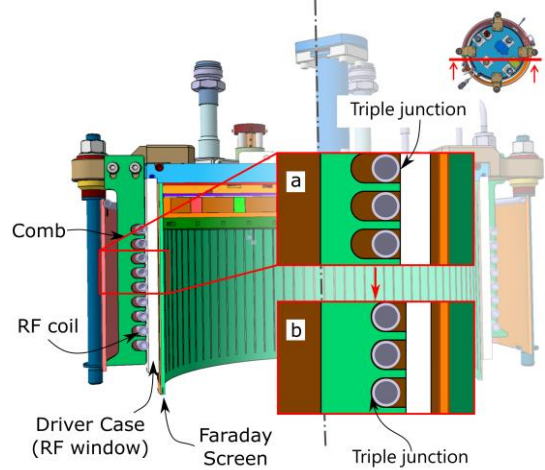

**Figure 9 – CAD view of a driver cross section in the original configuration, with a zoomed view of the RF coil touching the RF window generating a triple junction in the original configuration (a) and the improved configuration**

where the triple junction is at the contact point between the comb and the RF coil (b)

## 4. Magnetic filter field configuration improvement

In SPIDER a magnetic filter field is exploited for electron temperature and density reduction in the extraction region, and for filtering out the unavoidable co-extracted electrons. The filter field is created supplying with ISPG a current up to 5 kA on the PG and circulating it through a set of bus-bars.

The original design of the filter field was focused on the optimization of the B-field lines in the expansion volume and in the accelerator regions, minimizing it in the drivers [23]. Figure 10 a) sketches a cross section of the beam source components with red rectangles: a pair of RF drivers, the plasma expansion region, the PG, the EG and the GG; also the 5 original bus-bars feeding the PG indicated as "LEFT FWD", "RIGHT FWD", "LEFT RETURN", "MID RETURN", "RIGHT RETURN" are depicted.

During SPIDER operation when the PG current (magnetic field intensity) was increased beyond a certain threshold, the Hα reduced linearly up to a complete quench of the plasma. By analysing the B-field map, a configuration that caused electron loss on the driver walls close to the mid return bus-bar was identified.

The design of a new layout for the return bus-bars, capable of creating an optimized magnetic field topology within the RF driver, while keeping it almost unchanged in the other regions was studied [24]. In particular additional bus-bars, indicated as "NEW LEFT RETURN", "NEW MID RETURN" and "NEW RIGHT RETURN" in Figure 10 b), were placed on the backside of the source. As it can be seen in the same figures, the obtained B-field component axial to the RF driver (Bz) is relatively low with respect to the original configuration, which means that the field lines intercepting the driver walls are minimized.

This modification was introduced during a short SPIDER shutdown and proved being effective, since afterward the Hα grows with the increase of filter field current.

## 5. Fast transients due to grid breakdowns

Breakdowns across the grids can occur routinely during Neutral Beam Injector operation. The PS system was designed accordingly, implementing provisions to limit the arc energy.

With acceleration voltage higher than 50 kV, fast transient voltage peaks in the PS system components at the grids breakdowns were identified.

With reference to Figure 11, at the breakdown the current $I_A$ coming from the AGPS (and from the stray capacitance of the cable and HVD with respect to ground, not shown in the figure for simplicity) splits into $I_P$ that flows through the ISEG output filter toward the PG and $I_E$ that flow toward the EG. The $I_P$ causes an inverse voltage peak at the ISEG output filter that exceeds the ratings. The stray inductance of the AGPS output cable shield causes a voltage drop due to $I_A$, which exceeds the insulation level of the AGPS return path connections towards their grounded supports.

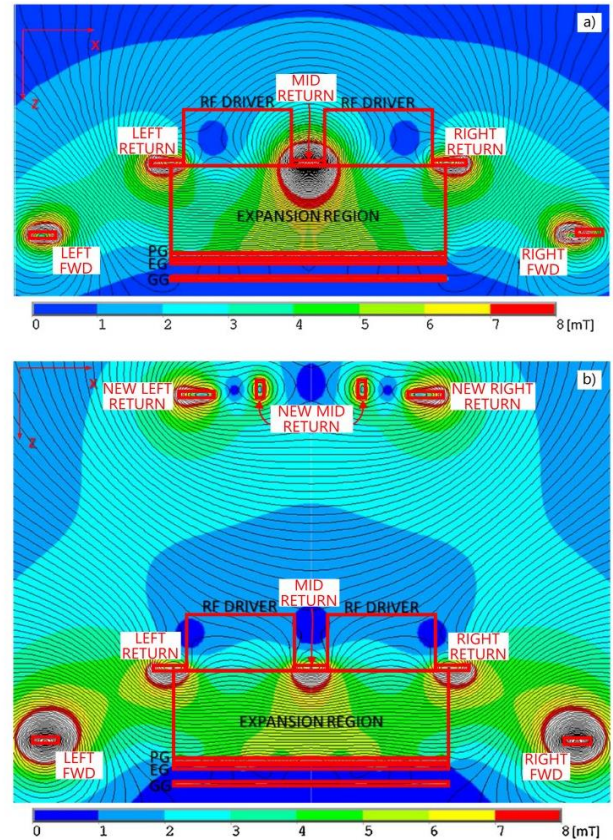

Figure 10 – Contour plot of the absolute value of the magnetic field: a) before and b) after the modification of the return bus-bars [24]

The circuit was modified as shown in Figure 12: ISEG output filter resistance was partialized for the $I_P$, while the AGPS cold pole was connected to ground through a passive RC network. The first modification proved being effective up to 80 kV as explained in [25], but it was possible to be implemented since ISEG output current was limited (because of the PG mask, see section 3). It cannot be thus a long-term solution: ISEG full current cannot flow through the partialized filter resistance because of voltage drop and dissipated power. As a long-term solution the redesign of the output filter and the adoption of suitable voltage clamps shall be considered.

The RC connected to the cold pole of AGPS proved being effective too, reducing the current carried by the cable screen and allowing current circulation through the grounding system.

As long-term solution, the insulation strength of the return path versus ground shall be improved, and it could be attempted to reduce the stray inductance of the cable shield with a better grounding of the cable tray [26].

Besides the described overvoltage issues, at the grid breakdowns the electronic cards implemented on some PS suffered for EMC issues that stop SPIDER pulses.

These electronic cards are custom boards used for the acquisition of the PS output voltage and current (analogue signals) and for the transmission through optic fibers. The EMC issue was due to fast voltage transients applied to the electronic boards between the ac voltage supply and the measuring point (physically connected or by means of the current probe stray capacitance). The voltage transient at the breakdown is due to the stray inductance of the cable connecting the reference potential to the ac supply, the issues on the boards are caused because the transient voltage is applied to the internal transformers and electronics. The solution, proved being effective up to the nominal SPIDER acceleration and extraction voltage values, was to reduce the stray inductance of the cable and to add suitable high-pass filters between the analogue inputs and the supply voltage reference [25].

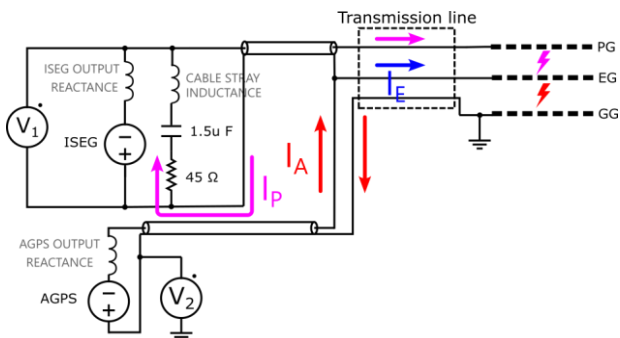

**Figure 11 – Simplified sketch of the path for the AGPS breakdown current**

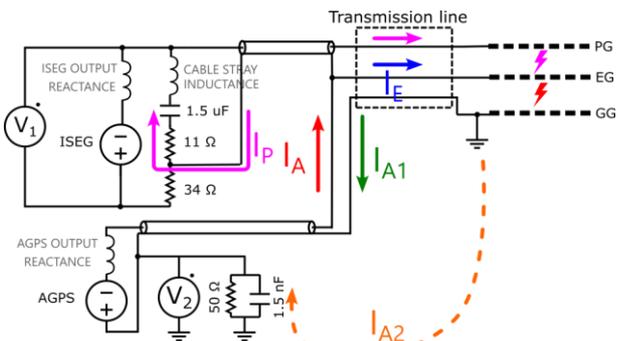

**Figure 12 – Simplified sketch of the electrical scheme of ISEG output filter and AGPS cold pole ground connection after the modification**

## 6. Conclusions

During 3 years of SPIDER operation we discovered many issues related to the extremely interconnected Power Supply system and the complex Beam Source.

Performing in parallel modelling activities and experimental campaigns allowed a deep understanding of the observed phenomena.

Short-term countermeasures installed during short shutdowns gave the possibility to verify the understanding and to estimate their effectiveness. The identified long-term solution are being designed in detail for implementation during the ongoing long shutdown.

The experience gained with SPIDER is of paramount importance in view of MITICA operation, where the 10 time higher acceleration voltage and the 3 times longer multi-conductor transmission line are expected being critical not only for the beam source but also for the PS system.

**Acknowledgments**

This work has been carried out within the framework of the ITER-RFX Neutral Beam Testing Facility (NBTF) Agreement and has received funding from the ITER Organization. The views and opinions expressed herein do not necessarily reflect those of the ITER.

This work has been carried out within the framework of the EUROfusion Consortium, funded by the European Union via the Euratom Research and Training Programme (Grant Agreement No 101052200 — EUROfusion). Views and opinions expressed are however those of the author(s) only and do not necessarily reflect those of the European Union or the European Commission. Neither the European Union nor the European Commission can be held responsible for them.